# Identifiability in Epidemic Models with Prior Immunity and Under-Reporting

Fanny Bergström[1] · Martina Favero[1] · Tom Britton[1]

**Abstract**

Identifiability is the property in mathematical modelling that determines if model parameters can be uniquely estimated. For infectious disease models, failure to ensure identifiability can lead to misleading parameter estimates and unreliable policy recommendations. We examine the identifiability of a modified Susceptible-Infectious-Recovered (SIR) model that accounts for under-reporting and pre-existing immunity in the population. We provide a mathematical proof of the structural unidentifiability of the deterministic model of jointly estimating three parameters: the fraction under-reporting, the proportion of the population with prior immunity, and the community transmission rate, when only reported case data are available. We then show, analytically and with a simulation study using a stochastic model, that the identifiability of all three parameters is achieved if the reported incidence is complemented with sample survey data of prior immunity or prevalence during the outbreak. Our results show the limitations of parameter inference in partially observed epidemics and the importance of identifiability analysis when developing and applying models for public health decision making.

## 1 Introduction

Identifiability is the concept in mathematical modelling that refers to the ability to uniquely infer model parameters. Identifiability can be divided into two categories: structural and practical identifiability (Wieland et al. 2021). Structural identifiability is the property that considers whether model parameters can be uniquely determined

✉ Fanny Bergström
fanny.bergstrom@math.su.se

Martina Favero
martina.favero@math.su.se

Tom Britton
tom.britton@math.su.se

[1] Department of Mathematics, Stockholm University, 106 91 Stockholm, Sweden

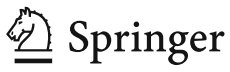

from the mathematical structure of the model. If this condition is not met, the model remains unidentifiable even with perfect data. Practical identifiability, on the other hand, considers which parameters are identifiable given the available data. While structural identifiability is necessary, it does not ensure unique parameter estimates in practice, as additional factors, such as noise or limited data, can affect the identifiability. Importantly, structural identifiability is a prerequisite for practical identifiability as model parameters cannot be uniquely estimated regardless of data quality without it. Moreover, structural identifiability analysis is essential in experimental design, as it helps identify which additional model outputs or measurements may be required to make a model identifiable.

Here, we focus on the structural identifiability of a modified Susceptible–Infectious–Recovered (SIR) model that incorporates two important factors for epidemics: under-reporting of cases and pre-existing immunity in the population. These are two common factors to consider for epidemics. For example, as COVID-19 has entered a phase with low testing and immunity in the population, this needs to be accounted for. Many infections now go unreported due to asymptomatic cases, which have been shown to contribute to transmission (Bai et al. 2020). Other factors contributing to the under-reporting are limited testing infrastructure and changing surveillance strategies. Simultaneously, varying degrees of immunity resulting from prior infections and vaccinations have altered population susceptibility, adding further complexities to the analysis. Failing to account for these factors in epidemic models can lead to significant biases in parameter estimates (Nguyen et al. 2023). For example, underestimating the basic reproduction number $R_0$ due to prior immunity or case under-reporting could result in delayed or insufficient public health interventions (Gibbons et al. 2014).

There are various approaches to structural identifiability analysis, including Laplace transform techniques (Bellman and Åström 1970), Taylor series expansion (Pohjanpalo 1978), and differential algebra approaches (Ljung and Glad 1994), the latter implemented in software such as DAISY (Bellu et al. 2007) and Julia (Dong et al. 2023). In this work, we use the Julia implementation and complement it with a direct reformulation of the ordinary differential equation (ODE) system into an explicit form that can be analysed directly. This reformulation is closely related to the Taylor?series approach of Pohjanpalo (1978) and is particularly useful here because it exposes the parameter relationships responsible for the unidentifiability. While software tools offer scalability, automation, and the ability to identify candidate parameter combinations, analytical approaches provide deeper insight into why unidentifiability occurs. They reveal the structural relationships among parameters, making symmetries, functional redundancies, and possible reparametrizations explicit. This clarifies the underlying sources of unidentifiability and supports more principled model reduction, parameter grouping, and experimental design.

Identifiability issues in compartmental epidemic models have received considerable attention (Evans et al. 2005; Kao and Eisenberg 2018; Magal and Webb 2018; Dankwa et al. 2022; Mielke and Christiansen 2023; Petrica and Popescu 2023). Still, the joint impact of under-reporting and pre-existing immunity has not been thoroughly investigated. Here, we provide mathematical proof of the structural unidentifiability within a deterministic SIR framework of three important epidemiological parameters: the reporting probability, the proportion of the population with pre-existing immunity, and the transmission rate. We then show that it is possible to resolve the unidentifiability and reliably infer two out of the three parameters by incorporating supplementary data from sample surveys, either as seroprevalence estimates or prevalence measurements. Furthermore, we use the structural identifiability of the deterministic model to shed light on the practical identifiability of the stochastic model with a simulation study. Our findings show the importance of identifiability analyses as a routine part of model development and application.

This paper is structured as follows. Section 2 introduces the mathematical definition of structural identifiability, the formulation of the SIR model and the simulation framework. Section 3 presents the theoretical results on structural unidentifiability and demonstrates how survey-based data can restore identifiability. Section 4 discusses the broader implications of our findings for inference and model-based decision making.

## 2 Method

### 2.1 Definition of Structural Identifiability

We consider a parametrized dynamical system

$$\begin{aligned} \dot{x}(t) &= f(x(t), \theta, u(t)), \\ y(t) &= h(x(t), \theta), \\ x_0 &= x(t_0, \theta) \end{aligned} \tag{1}$$

with vector functions $f$ and $h$, parameter vector $\theta \in \Theta$, state space vector $x(t)$, known input vector $u(t)$, and observed output $y(t)$ (Villaverde et al. 2016). For a model specified as in (1), a parameter $\theta_i$ is structurally globally identifiable if

$$\forall \theta, \theta' \in \Theta, \qquad y(t, \theta) = y(t, \theta') \text{ for all valid inputs } u(t) \implies \theta = \theta'.$$

The definition of local structural identifiability is similar but instead of the complete parameter space, only some neighbourhood of $\Theta$ is considered. If neither global nor local identifiability holds, the parameter is structurally unidentifiable. An unidentifiable parameter $\theta_i$ means that a change in this parameter can be completely compensated by a change in another parameter, resulting in two identical model outputs. The model is structurally identifiable only if all parameters in $\theta$ are structurally identifiable.

### 2.2 Stochastic SIR Model

We consider a stochastic SIR model to describe the dynamics of an epidemic in a closed population of known constant size $n$, with homogeneous mixing. That is, every individual in the population is equally likely to come into contact with any other individual. The model incorporates the possibility of under-reporting by distinguishing between reported and unreported infections.

A fraction $p \in [0, 1]$ of infectious individuals is reported, and the remaining fraction $1 - p$ is not reported. Let $S(t)$ denote the number of susceptible individuals at time $t$, and let $I_r(t)$ and $I_u(t)$ represent the numbers of reported and unreported infectious individuals, respectively. Similarly, let $R_r(t)$ and $R_u(t)$ denote the numbers of recovered individuals who were reported or unreported while infectious. The number of individuals in each compartment at time $t$ are counts ranging from 0 to $n$. A proportion $i_0 \in [0, 1]$ of the population is initially reported infectious, i.e., there are $ni_0$ number of reported infections at time $t = 0$ (and $\frac{1-p}{p} ni_0$ unreported), where we assume $i_0$ to be small as we consider this to be at an early stage of the outbreak. Another proportion $\pi = (R_r(0) + R_u(0))/n \in [0, 1]$ is assumed to be initially immune due to prior exposure or vaccination. Infectees have infectious contacts with susceptible individuals at different rates, specifically reported individuals make contact at a rate $\beta_r \in (0, \infty)$ and unreported individuals at a rate $\beta_u \in (0, \infty)$ with a random individual in the population with uniform probability $1/n$. An infectious individual recovers after a duration that follows an exponential distribution with rate $\gamma \in (0, \infty)$, corresponding to an average infectious period of $1/\gamma$. Upon recovery, individuals move to $R_r(t)$ or $R_u(t)$, depending on whether they were reported or unreported.

To obtain realisations of the epidemic trajectories, we simulate the model described above using the Gillespie algorithm (Gillespie 1976). The algorithm is implemented in R (R Core Team 2025, v4.5.1) and shown in Appendix A.

### 2.3 The Deterministic Model

In the large population limit ($n \to \infty$) the stochastic SIR model is equal to the deterministic model (Diekmann 1977). We can write the ODEs of the deterministic SIR model with under-reporting as

$$
\begin{aligned}
\frac{d}{dt}S(t) &= -\left(\beta_r I_r(t) + \beta_u I_u(t)\right)\frac{1}{n}S(t) \\
\frac{d}{dt}I_r(t) &= p\left(\beta_r I_r(t) + \beta_u I_u(t)\right)\frac{1}{n}S(t) - \gamma I_r(t) \\
\frac{d}{dt}I_u(t) &= (1-p)\left(\beta_r I_r(t) + \beta_u I_u(t)\right)\frac{1}{n}S(t) - \gamma I_u(t) \\
\frac{d}{dt}R_r(t) &= \gamma I_r(t) \\
\frac{d}{dt}R_u(t) &= \gamma I_u(t),
\end{aligned} \tag{2}
$$

with $S(t) + I_r(t) + I_u(t) + R_r(t) + R_u(t) = n$ at all times $t$. We use initial conditions

$$
\begin{aligned}
S(0) &= n - I_r(t) - I_u(t) - R_r(t) - R_u(t) = n(1-\pi) - n\frac{1}{p}i_0 \\
I_r(0) &= ni_0 \\
I_u(0) &= n\frac{1-p}{p}i_0 \\
R_r(0) &= np\pi \\
R_u(0) &= n(1-p)\pi.
\end{aligned} \tag{3}
$$

Since the identity of the infector is generally unobserved in the data, disentangling $\beta_r$ and $\beta_u$ based solely on incidence data is infeasible. To account for this, we define an effective infection rate

$$\beta^* = p\beta_r + (1-p)\beta_u. \tag{4}$$

This parameter captures the weighted average contribution to transmission from both reported and unreported cases, allowing inference from incidence data. If we let $I = I_r + I_u$ and $R = R_r + R_u$ this model is equal to the standard SIR model with $\beta^*$ as infection rate (Kermack and Mckendrick 1927).

For our model, we have the following expression for the basic reproduction number $R_0 = \beta^*/\gamma$ and the effective reproduction number (taking into account prior immunity) $R_E = \beta^*(1-\pi)/\gamma$. From the deterministic model we have two equations: the initial growth rate and the final size of the epidemic (Diekmann et al. 2013). The initial exponential growth rate $\rho$ (of incidence, number infected and currently infectious individuals) equals

$$\rho = \beta^*(1-\pi) - \gamma. \tag{5}$$

If we let $I = I_u + I_r$, then

$$\frac{d}{dt}I(0) = \frac{1}{p}ni_0\left[\beta^*(1-\pi) - \gamma -' \frac{\beta^*}{p}i_0\right],$$

where the first two terms within the brackets correspond to the initial growth rate.

The final fraction $z$ of infected during an outbreak, will in a large population lie close to the solution to the equation

$$1 - z = e^{-R_E z}. \tag{6}$$

We let $z_r$ denote the final fraction of reported infections in the whole population such that $z_r = p(1-\pi)z$. Similar to (6), the final fraction reported infections $z_r$ will be close to the solution of the equation

$$1 - \frac{z_r}{p(1-\pi)} = e^{-R_E \frac{z_r}{p(1-\pi)}}. \tag{7}$$

Equations (5) and (7) are used in Section 3.2 to estimate parameters $p, \pi$ and $\beta^*$ from simulated data.

## 2.4 Inference

The modified stochastic SIR model defined by the ODEs in (2) can also be defined by four counting processes $N_1$, $N_2$, $N_3$, and $N_4$. $N_1$ and $N_2$ count the number of reported infections and recoveries and $N_3$ and $N_4$ the unreported infections and recoveries, respectively. We can write the probability of a specific realization of the epidemic in the time interval $[0, t)$ using the counting processes and martingale theory (Diekmann et al. 2013, p. 138). These likelihood and corresponding log-likelihood is presented in

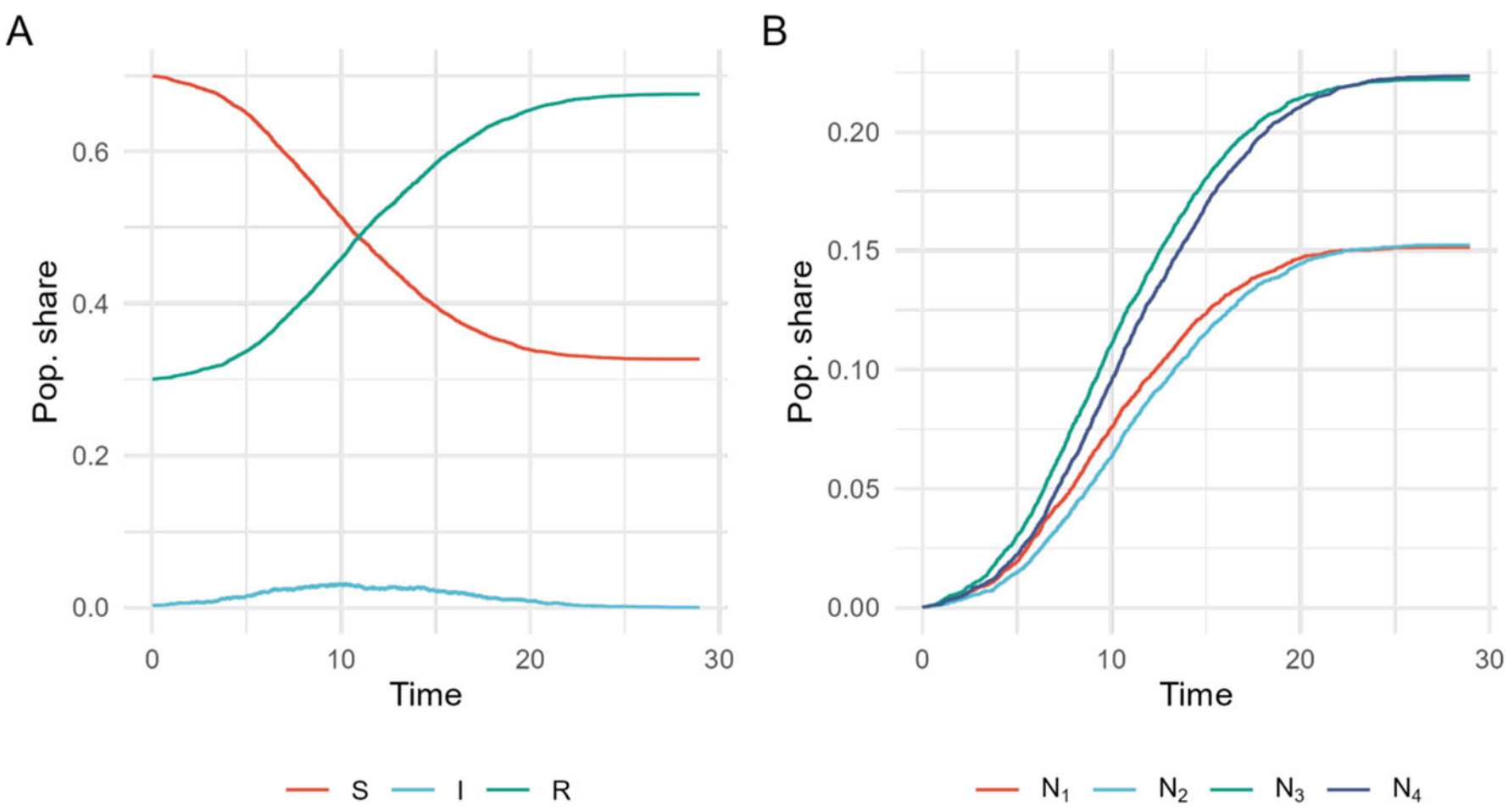

**Fig. 1** A simulated epidemic of the SIR model with under-reporting. (**A**) The compartments (joint reported and unreported) of the susceptible ($S$), infectious ($I = I_r + I_u$) and recovered ($R = R_r + R_u$) individuals over time. (**B**) The counting processes of new infections reported and unreported ($N_1$ and $N_3$) and their respective recoveries ($N_2$ and $N_4$). The counting process of the reported infections, $N_1$, is the observable process that we consider as the data in this analysis

Appendix B. The log-likelihood expressed in (10) assumes that we can observe everything (infection times and recovery times for all individuals, including unreported). However, in our analysis we are interested in inference that can be based on reported incidence. Hence, instead of $\beta_r$ and $\beta_u$, we consider the effective transmission rate $\beta^*$ defined in (4) to be the parameter of interest. We also consider the recovery rate $\gamma$ to be estimated separately and here assumed to be a known quantity. For the case when we assume that we do not observe infection and recovery time, there is no simple explicit likelihood. An approximation could be made by replacing unreported by reported multiplied by $(1 - p)/p$. We could also rewrite $S(t)$ in terms of the counting process $N_1$, i.e.

$$S(t) = n(1 - \pi) - \frac{1}{p} N_1(t), \tag{8}$$

to introduce $\pi$ into the likelihood. However, the approximate likelihood (equation (11), Appendix B) cannot be used to infer $\beta^*$, $p$ and $\pi$. Instead, we will use the limiting equations (5) and (7) for the inference in our analysis.

A stochastic SIR epidemic model is simulated using the Gillespie algorithm as described in Section 2.2. A simulated outbreak is seen in Figure 1 with parameter values found in Table 1. We have not given the rate parameters time units, e.g. days or weeks, instead we measure the time with respect to an average infectious periods as we set $\gamma = 1$, implying that $\beta * = R_0$. For example, one infectious period can be interpreted as one week, similar to that of an average influenza-like infection (Jing et al. 2020). The basic reproduction number is set to $R_0 = 1.9$, also corresponding to a highly transmissible influenza-like infection (Biggerstaff et al. 2014). Furthermore, we assume that 30% of the population has prior immunity at the start of the epidemic and a reporting fraction of 40%. These assumptions allow us to simulate a scenario representative of a seasonal influenza outbreak. The infection rate for reported and unreported infections is in our simulation given separate values. We do not make any explicit assumption about the one being higher than the other and believe this could go in either direction. Individuals who report their infections might have more severe symptoms and therefore isolating themselves more seriously, on the other hand unreported may have less severe symptoms so they also transmit the disease to a smaller extent.

**Table 1** Parameter and initial values for the simulation of the SIR model with under-reporting.

| Variable | Notation | Value |
|---|---|---|
| Population size | $n$ | 10,000 |
| Initially infectious (reported) | $ni_0$ | 10 |
| Initially infectious (unreported) | $ni_u$ | 15 |
| Infection rate (reported) | $\beta_r$ | 2.5 |
| Infection rate (unreported) | $\beta_u$ | 1.5 |
| Infection rate (overall) | $\beta^* = R_0$ | 1.9 |
| Reporting fraction | $p$ | 0.40 |
| Fraction initially immune | $\pi$ | 0.30 |
| Recovery rate | $\gamma$ | 1.00 |
| Exponential growth rate | $\rho$ | 0.33 |
| Fraction reported infected during outbreak | $z_r$ | 0.13 |

Classical compartmental transmission models assume exponential growth during the early phase of a well-mixed population (Diekmann et al. 2013, p. 10). We use this assumption to estimate the exponential growth rate $\rho$ of our simulated epidemic. In practice, we will fit a linear model to the log of the cumulative number of reported infections $N_1(t)$. Because exponential growth implies

$$N_1(t) \approx N_1(0)e^{\rho t},$$

taking logarithms yields a linear a linear relationship with slope $\rho$. We therefore regress $\log N_1(t)$ on time using ordinary least squares, minimising the discrepancy between the linear prediction and the observed log?cumulative incidence. Although one might expect earlier time points to be more informative about the initial growth rate, in practice these observations are highly stochastic, therefore provide less stable estimates. Using each infection time as a data point gives more weight to periods with more infections, reducing the influence of early stochastic fluctuations and produces a more reliable estimate of $\rho$. We consider the initial stage of the epidemic to last until 7.5% of the population has been reported infected.

## 3 Results

### 3.1 Unidentifiability

We consider the deterministic SIR model with under-reporting described in Section 2.3. The model depends on three unknown parameters: the infection rate $\beta^*$, the under-reporting fraction $p$, and the fraction initially immune $\pi$. The two features of the incidence curve that turn out to carry information about model parameters is the initial growth rate $\rho$ and the cumulative reported incidence, i.e. the reported final epidemic size $z_r$. Table 3 in Appendix C provides a complete list of the variables and parameters used in our analysis.

Using the Julia package `StructuralIdentifiability.jl` (Dong et al. 2023), we verified that the model is structurally unidentifiable when only reported infections are observed. In particular, multiple parameter triples $(\beta^*, p, \pi)$ are consistent with the same observed growth rate $\rho$ and final reported size $z_r$. This is illustrated with a simulated epidemic in Figure 1. From the simulated reported incidence we estimate the exponential growth rate is $\hat{\rho} = 0.346$ (95% CI: 0.344–0.349), close to the true value 0.333. Together with the observed final fraction reported cases $z_r$, a range of different triples $(p, \pi, \beta^*)$ produce identical values of $(\rho, z_r)$, as shown in Figure 2B. The apparent excess growth in the very early phase of the simulation arises because the epidemic is initiated with 25 fully infectious index cases. Had the outbreak started with a single index case, only about 10–15 individuals would still be infectious by the time 25 cumulative infections had occurred. Thus, the earliest part of the simulated trajectory is atypical and should not be interpreted as reflecting the underlying exponential growth rate.

For our application we are interested in deriving the identifiable relationship between the three unknown parameters, which is not learned through the software. To obtain it, we apply the analytic method described in Appendix D, which yields an explicit characterization of the identifiable parameter relationship and forms the basis for the results that follow.

**Theorem 1** (Structural Unidentifiablity) *Let $I_r(t, \theta)$ be a trajectory representing the numbers of reported infectious individuals driven by the ODEs of* (2) *and* (3) *with an effective infection rate $\beta^*$ defined in* (4) *for a parameter set $\theta = (\pi, \beta^*, p)$.*

*If $I_r(t, \theta_1) = I_r(t, \theta_2)$, it must be that*

$$\frac{\beta_1^*}{p_1} = \frac{\beta_2^*}{p_2} \quad \text{and} \quad \beta_1^*(1 - \pi_1) = \beta_2^*(1 - \pi_2).$$

*Vice versa, if the two sets of parameters satisfy the conditions in the display above, and the number of reported initially infectious individuals is the same, then the two trajectories must be identical.*

**Corollary 2** (Identifiability) *Assume that one of the three parameters is known (previously estimated). If the two trajectories are identical, then it must be that $\theta_1 = \theta_2$.*

The proof of Theorem 1 is provided in Appendix D. Figure 3 shows an example of two sets of parameter values that gives an identical trajectory of the reported cases

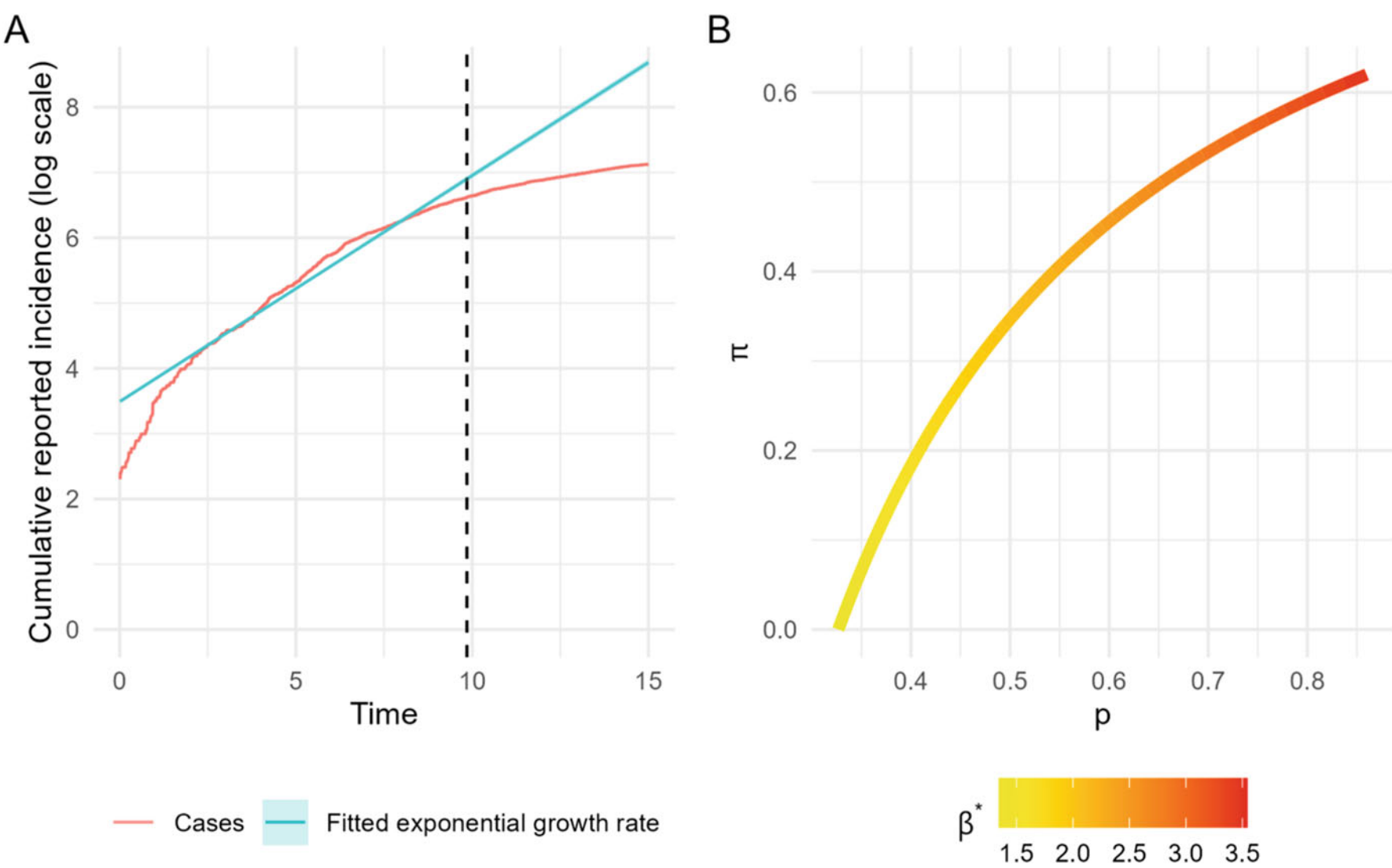

**Fig. 2** (**A**) Reported cumulative incidence (red) on the log-scale and estimated exponential growth rate (blue) achieved by fitting a linear model to the reported incidence. The shaded area (largely indistinguishable due to its narrow width) is the 95% confidence interval of the fitted model. The dashed vertical line marks where 7.5% of the population has had a reported infection, which is the end of the fitting period of the linear model. (**B**) The combinations of possible values of $p, \pi$ and $\beta^*$ that give the same value of the growth rate and final size of reported cases (Color figure online)

for the deterministic model. The values of the first parameter set are the values found in Table 1. The second set of parameter values are chosen so that $\pi = 0$ and that they satisfy $\frac{\beta_1^*}{p_1} = \frac{\beta_2^*}{p_2}$ and $\beta_1^*(1 - \pi_1) = \beta_2^*(1 - \pi_2)$, which gives us $p = 0.24$ and $\beta^* = 1.14$.

Beyond the structural unidentifiability, the figure further illustrates that ignoring prior immunity leads to an underestimation of $\beta^* = R_0 = 1.14$ compared to the true value 1.9 when ignoring the prior immunity $\pi = 0.3$. The implication of the shown unidentifiability is that we need to estimate $p$ or $\pi$ (or both) from other data to infer the three parameters, and to infer $R_0$. However, if we know the fraction initially immune $\pi$ this also means that we can estimate the reporting fraction $p$ by only observing the reported incidence over time. Likewise, if we know $p$, we can uniquely infer the remaining two parameters $\beta^*$ and $\pi$.

### 3.2 Estimation From Data

As concluded in Section 3.1, the parameters $p, \pi$ and $\beta^*$ cannot be estimated simultaneously because of the shown structural unidentifiability, but if information about either $p$ or $\pi$ is available from other sources we can estimate the remaining two parameters. Recall that $\gamma$ is considered known (estimated separately). We use simulations of 100 epidemics resulting in major outbreaks with the parameter values in Table 1. We consider two scenarios in which we can obtain sample survey data to estimate $\pi$

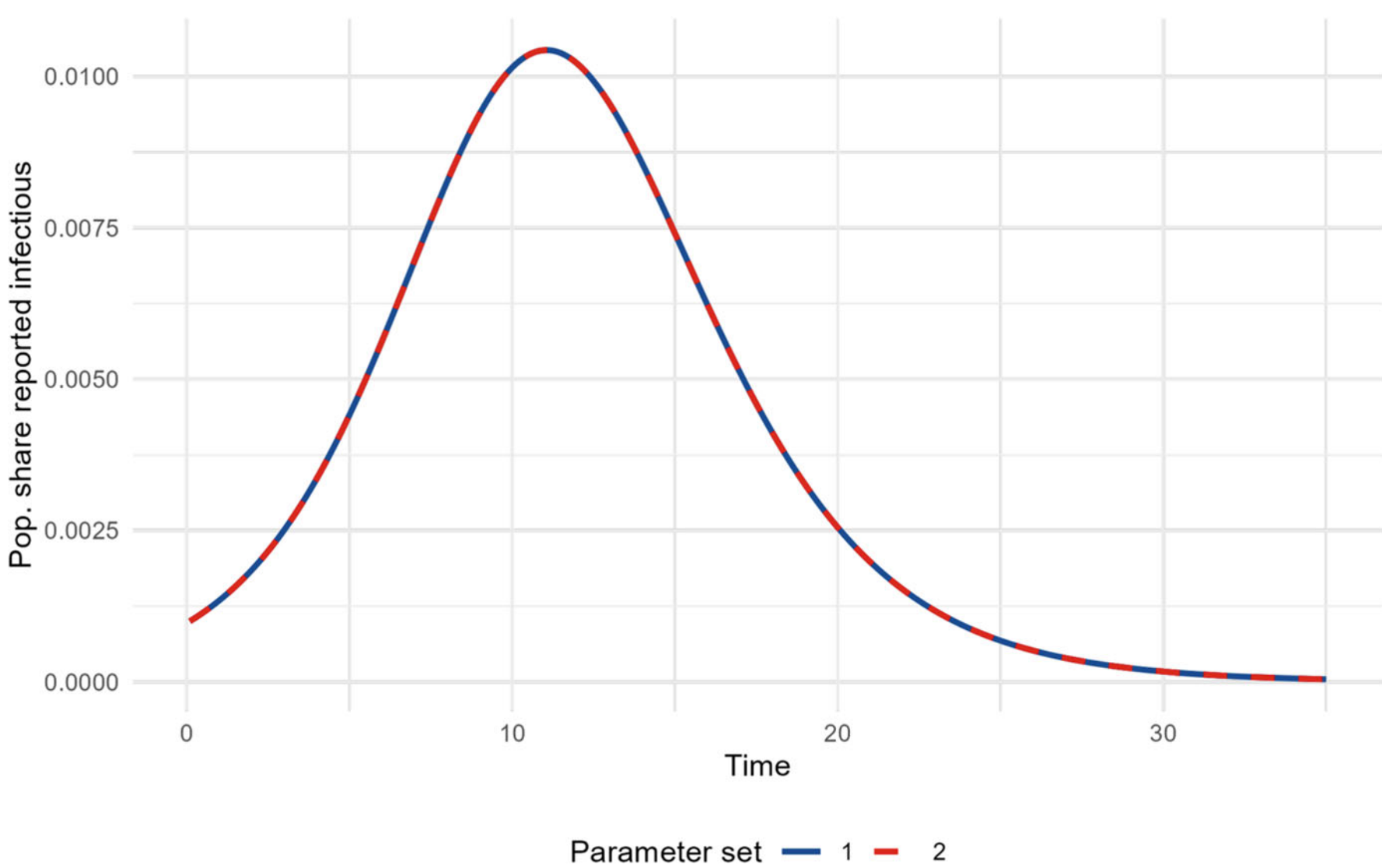

**Fig. 3** Illustration theorem 1, showing two sets of parameters resulting in identical trajectories of the reported infectious individuals $I_r(t)$ in the deterministic SIR model. The sets of parameters are $\{p_1 = 0.4, \pi_1 = 0.3, \beta_1^* = 1.9\}$ and $\{p_2 = 0.24, \pi_2 = 0.0, \beta_2^* = 1.14\}$ (Color figure online)

or $p$. In both scenarios, we sample 1,000 individuals at random without replacement from the population. We let $\hat{\pi}$ be the fraction immune within the sample taken at time $t = 0$ and $\hat{p}$ be the fraction reported among infectious individuals at the peak of the epidemic.

For each of the 100 simulated epidemics, we estimate the initial exponential growth rate $\hat{\rho}$ as described in Section 2.3. For each simulation there is also an uncertainty in the estimation of the linear model but here we disregard this as the larger variation will come from the differences in the simulated outbreak. For each simulation, we also get a fraction $\hat{z}_r$ of the final size of the reported epidemic. With the quantities $\hat{\rho}$ (mean: 0.312, SD: 0.055) and $\hat{z}_r$ (mean = 0.123, SD = 0.010) from the simulation and data on $\hat{p}$ or $\hat{\pi}$ from a sample of the population, we can now estimate the remaining two parameters using the two equations (5) and (7) from the deterministic model. Although one could apply an optimisation procedure to identify the last two parameters that best match the data, this is not required because the two equations are sufficient to obtain the parameters directly. The mean and their respective standard deviation of the estimated parameters from the 100 simulations are shown in Table 2. The parameters $p$ and $\pi$ are fractions ranging from 0 to 1 and $\beta^* = R_0$ is dimensionless in the sense that we measure time in infectious periods as $\gamma = 1$.

The results show that we are able to estimate two of the three parameters with an estimate of $\pi$ or $p$. In our case, with large pre-existing immunity (30%) we can obtain an estimate $\hat{\pi}$ that is close to the true value with a small variance, from which we can derive the remaining two unknown parameters with good precision. Estimating $p$ at the peak of the epidemic is a harder task since the number of infectious individuals at this stage is lower compared to the number of individuals with pre-existing immunity.

**Table 2** Mean results from estimating parameters from 100 simulations resulting in major outbreaks for the stochastic SIR model with under-reporting in a population of size $n = 10,000$. We assume that we can estimate either $\hat{\pi}$ or $\hat{p}$ from a random sample of the population from which we can estimate the remaining two parameters.

| Parameter | True value | Estimate with $\hat{\pi}$ (standard deviation) | Estimate with $\hat{p}$ (standard deviation) |
|---|---|---|---|
| $p$ | 0.4 | 0.426 (0.041) | 0.407 (0.028) |
| $\pi$ | 0.3 | 0.299 (0.015) | 0.265 (0.088) |
| $\beta^* = R_0$ | 1.9 | 1.875 (0.091) | 1.817 (0.286) |

This leads to an estimate of $\beta^* = R_0$ with a larger standard deviation when inferred from the sample survey data of $\hat{p}$ compared to $\hat{\pi}$.

## 4 Discussion

We explored the identifiability of an SIR model accounting for under-reporting and prior immunity. The structural identifiability results obtained from software-based tools are strengthened by our analytical approach by revealing the relationships and symmetries that generate indistinguishable epidemic trajectories. In doing so, our analysis clarifies the underlying structural constraints and highlights the parameter dependencies, making it a complement to computational identifiability methods.

We show that the three key parameters of an epidemic, a transmission rate $\beta^*$, the reporting fraction $p$, and the immune fraction at the beginning of the epidemic $\pi$, are not structurally identifiable based solely on the reported incidence over time. Multiple combinations of parameters can reproduce the same epidemic curve of reported incidence, while implying different values of the basic reproduction number $R_0$. However, our analysis also shows that a consistent estimation of all three parameters becomes feasible from reported incidence together with sample survey data, either measuring prior immunity or estimating infection prevalence. Hence, we showed that if one of the three parameters is known, the three parameters become identifiable and can be derived from the equations of the exponential growth rate (5) and final size of the epidemic (7).

A limitation of our analysis is the assumption of a fixed reporting fraction $p$ over time. In reality, this is likely to change as testing strategies or healthcare-seeking behaviour change. To overcome this problem, multiple prevalence samples during the outbreak is probably necessary. Furthermore, more complex transmission models, such as Susceptible-Exposed-Infectious-Recovered (SEIR) or Susceptible-Infectious-Recovered-Susceptible (SIRS) models, could also be considered. The SEIR models better capture realistic disease progression by including a latent stage. However, their use also introduces additional complications, particularly the need to know or estimate the generation time distribution. The exponential growth rate of the epidemic depends on this distribution via the Euler-Lotka equation, meaning that incorrect assumptions here can lead to significant biases. Incorporating temporary immunity in an SIRS model may also provide a more realistic representation for diseases such as seasonal influenza or COVID-19. Yet this, too, increases analytical complexity, as the standard SIR model and an SIRS model with temporary immunity are structurally indistinguishable when only the infectious class is observed (Evans et al. 2004).

Our work underscores the importance of accounting for prior immunity. In many scenarios, a non-negligible proportion of the population may have immunity at the onset, whether due to previous exposure or vaccination. Ignoring this can result in a substantial underestimation of $R_0$, as shown in Figure 3. If under-reporting and prior immunity are present but not accounted for, even well-fitting models can yield misleading conclusions.

We used a simplified SIR model with homogeneous mixing and no interventions to isolate the identifiability problem for the three chosen parameters: $p$, $\pi$ and $\beta^* = R_0$. For instance, the infectious period $\gamma$ was not inferred in our analysis, but was assumed to be known or estimated independently, for example from clinical data. We also assume a known fixed population of size $n$, a more realistic setting would be to include birth and death processes and consider the population size as time-varying, and perhaps not known quantity as well. An exploration of how qualitative conclusions are affected not only by the population size $n$, but also on the true parameter values, in particular $R_0$, $p$ and $\pi$, would be an interesting follow up study. Although our model omits several complexities that would make it more realistic, such as age structure and behavioural changes, it allows for insights into the limitations of inference from reported incidence data alone. We also assume that the full epidemic is observed, but a potential further investigation is to what extent inference could be made at earlier time points of the epidemic, such as at its peak.

Finally, another direction for future work is the application to real data. Doing so would introduce practical challenges that are abstracted away in our theoretical analysis, as surveillance data are often incomplete or noisy. Also, assumptions of homogeneous mixing or constant reporting rates may not hold when analysing real data, and accounting for these deviations becomes essential for reliable inference. Applying the framework to real data would therefore require estimating quantities that were disregarded in the theoretical analysis, such as reporting delays, time-varying reporting probabilities, generation-interval distributions, or mixing patterns using external empirical studies or additional datasets. This would show further limitations and possibilities of parameter inference and help inform the design of surveillance systems and data collection strategies needed to support an effective public health response.

**Supplementary information** Appendix A contains the Gillespie algorithm used for the simulation of the stochastic modified SIR model with under-reporting. Appendix B presents the likelihood and approximated likelihood of the observable parameters $\beta^*$, $p$ and $\pi$. Appendix C provides a table with the data and parameters used in our analysis. Appendix D provides the mathematical proof of Theorem 1.

## Appendix A Gillespie Algorithm

We use the Gillespie algorithm (Gillespie 1976) to simulate a stochastic SIR model as specified in Section 2.2. The algorithm is outlined in Algorithm 1. This algorithm simulates the stochastic spread of an infectious disease using the modified SIR model with reported and unreported cases. It begins by initializing model parameters, the population states, and counters for tracking the counting processes. The simulation proceeds in continuous time by modelling each event, infection or recovery, as occurring at random intervals based on their respective rates. At each step, the force of infection is computed from the current number of infectious individuals, weighted by the transmission rates for reported and unreported cases (line 6). Using this force, the algorithm calculates the rates of four possible events: a reported infection, an unreported infection, a recovery from a reported case, or a recovery from an unreported case (line 7). The time until the next event is drawn from an exponential distribution and the specific event is chosen in proportion to its rate (line 9). Once an event is selected, the population state is updated accordingly and the new state is recorded (line 22). This process repeats until a predefined end time is reached or there are no infectious individuals remaining (line 23). The algorithm returns the complete time series of all the compartment values and the event counts (line 24).

**Algorithm 1** Simulation of the modified stochastic SIR model with under-reporting using the Gillespie Algorithm

1: **Input:** Parameters `params` $(\beta_r, \beta_u, p, \pi, \gamma)$, initial state `initial`, population size `pop`, end time `end_time`
2: Initialize state variables from `initial`: $S, I_r, I_u, R_r, R_u, N_1, N_2, N_3, N_4$
3: Initialize result list
4: $time \leftarrow 0$
5: **while** $time < end_time$ **and** $(I_r + I_u > 0)$ **do**
6: Compute infection force: $\lambda \leftarrow \beta_r I_r + \beta_u I_u$
7: Compute rates:

- $a_1 \leftarrow p\lambda S/pop$ ▷ Reported infection
- $a_2 \leftarrow (1-p)\lambda S/pop$ ▷ Unreported infection
- $a_3 \leftarrow \gamma I_r$ ▷ Reported recovery
- $a_4 \leftarrow \gamma I_u$ ▷ Unreported recovery

8: Total rate: $a_{total} \leftarrow a_1 + a_2 + a_3 + a_4$
9: Sample time to next event: $\tau \sim \text{Exponential}(a_{total})$
10: $time \leftarrow time + \tau$
11: Draw random number $t \sim \text{Uniform}(0, a_{total})$
12: Determine which event occurs:
13: **if** $t < a_1$ **then**
14: $S \leftarrow S - 1, I_r \leftarrow I_r + 1, N_1 \leftarrow N_1 + 1$
15: **else if** $t < a_1 + a_2$ **then**
16: $S \leftarrow S - 1, I_u \leftarrow I_u + 1, N_3 \leftarrow N_3 + 1$
17: **else if** $t < a_1 + a_2 + a_3$ **then**
18: $I_r \leftarrow I_r - 1, R_r \leftarrow R_r + 1, N_2 \leftarrow N_2 + 1$
19: **else**
20: $I_u \leftarrow I_u - 1, R_u \leftarrow R_u + 1, N_4 \leftarrow N_4 + 1$
21: **end if**
22: Record current state to result lists
23: **end while**
24: **return** Result list of $S, I_r, I_u, R_r, R_u, N_1, N_2, N_3, N_4$ over time

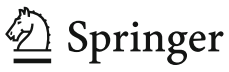

## Appendix B The Likelihood and Approximated Likelihood

We let $t_i$ denote the observed infection times, $\tau_i$ the unobserved infection times, and $f_\gamma(\cdot)$ the probability distribution function for the recovery time with parameter $\gamma$. We denote $\bar{S} = S/n$. Assuming $\beta_r$ and $\beta_u$ as well as the four counting processes $N_1$, $N_2$, $N_3$ and $N_4$ are observable, we can write the likelihood for the SIR model with under-reporting as

$$\begin{aligned} L_t(\beta_r, \beta_u, p, \pi, \gamma) &= p^{N_1(t)}(1-p)^{N_3(t)} \\ &\cdot \prod_{t_i} \big(\beta_r I_r(t_{i-}) + \beta_u I_u(t_{i-})\big)\bar{S}(t_{i-}) \cdot \prod_{\tau_i} \big(\beta_r I_r(\tau_{i-}) + \beta_u I_u(\tau_{i-})\big)\bar{S}(\tau_{i-}) \\ &\cdot e^{-\int_0^t (\beta_r I_r(s) + \beta_u I_u(s))\bar{S}(s)ds} \cdot \prod_{t_i} f_\gamma(t_i) \cdot \prod_{\tau_i} f_\gamma(\tau_i), \end{aligned} \tag{9}$$

where $t_{i-}$ denotes the time just before time $t_i$. Now assuming $\gamma$ to be known, the last two terms of (9) become two constants. We then write the corresponding log-likelihood as

$$\begin{aligned} l_t(\beta_r, \beta_u, p, \pi) \propto\ & N_1(t)\log(p) + N_3(t)\log(1-p) \\ & + \sum_{t_i} \log\big(\big(\beta_r I_r(t_{i-}) + \beta_u I_u(t_{i-})\big)\bar{S}(t_{i-})\big) \\ & + \sum_{\tau_i} \log\big(\big(\beta_r I_r(\tau_{i-}) + \beta_u I_u(\tau_{i-})\big)\bar{S}(\tau_{i-})\big) \\ & - \int_0^t (\beta_r I_r(s) + \beta_u I_u(s))\bar{S}(s)ds. \end{aligned} \tag{10}$$

To express the log-likelihood defined in (10) in terms of observable quantities, we use the $\beta^* = p\beta_r + (1-p)\beta_u$ and the approximations $N_3 \approx \frac{1-p}{p}N_1$ and $I_u \approx \frac{1-p}{p}I_r$. We also rewrite $S(t)$ in terms of the counting process $N_1$ and $\pi$ given (8), that is, $S(t) = n(1-\pi) - \frac{1}{p}N_1(t)$. We can express the log-likelihood for parameters $\beta^*$, $p$ and $\pi$ as

$$\begin{aligned} l_t(\beta^*, p, \pi) \propto\ & N_1(t)\log(p) + \frac{1-p}{p}N_1(t)\log(1-p) + \sum_{t_i}\log\big(\beta^*\frac{1}{p}I_r(t_{i-})\bar{S}(t_{i-})\big) \\ & + \frac{1-p}{p}\sum_{t_i}\log\big(\beta^*\frac{1}{p}I_r(t_{i-})\bar{S}(t_{i-})\big) - \int_0^t \beta^*\frac{1}{p}I_r(s)\bar{S}(s)ds \\ \propto\ & N_1(t)\log(p) + \frac{1-p}{p}N_1(t)\log(1-p) \\ & + \frac{1}{p}\sum_{t_i}\log\big(\beta^*\frac{1}{p}I_r(t_{i-})\bar{S}(t_{i-})\big) - \int_0^t \beta^*\frac{1}{p}I_r(s)\bar{S}(s)ds \qquad (11) \\ \propto\ & N_1(t)\log(p) + \frac{1-p}{p}N_1(t)\log(1-p) \end{aligned}$$

$$+\frac{1}{p}\sum_{t_i}\log\left(\beta^*\frac{1}{p}I_r(t_{i-})\big((1-\pi)-\frac{1}{p}\bar{N}_1(t_{i-})\big)\right)$$
$$-\int_0^t\beta^*\frac{1}{p}I_r(s)\big((1-\pi)-\frac{1}{p}\bar{N}_1(s)\big)ds.$$

We take the partial derivative of (11) with respect to $\beta^*$, $\pi$ and $p$.

$$\frac{\partial l_t}{\partial\beta^*}=\frac{1}{p}\frac{N_1(t)}{\beta^*}-\int_0^t\frac{1}{p}I_r(s)\big((1-\pi)-\frac{1}{p}\bar{N}_1(s)\big)ds$$
$$\frac{\partial l_t}{\partial\pi}=-\sum_{t_i}\frac{1}{p((1-\pi)-\frac{1}{p}\bar{N}_1(t_{i-}))}+\int_0^t\frac{\beta^*}{p}I_r(s)ds$$
$$\frac{\partial l_t}{\partial p}=-\frac{N_1(t)\log(1-p)}{p^2}-\frac{1}{p^2}\sum_{t_i}\log\left(\beta^*\frac{1}{p}I_r(t_{i-})\big((1-\pi)-\frac{1}{p}\bar{N}_1(t_{i-})\big)\right)$$
$$+\frac{1}{p^2}\sum_{t_i}\frac{\bar{N}_1(t_{i-})}{p(1-\pi)-\bar{N}_1(t_{i-})}-\frac{1}{p^2}N_1(t)$$
$$+\int_0^t\beta^*\frac{1}{p^2}I_r(s)\big((1-\pi)-\frac{2}{p}\bar{N}_1(s)\big)ds.$$

We can find an analytical solution for $\beta^*$. For the other two parameters we can find the solution numerically. If we equate $\frac{\partial l_t}{\partial\beta^*}$ to 0 and solving for $\beta^*$, we get the ML estimate

$$\hat{\beta}^*=\frac{N_1(t)}{\int_0^t I_r(s)\big((1-\pi)-\frac{1}{p}\bar{N}_1(s)\big)ds}.$$

We take the second derivatives of (11)

$$\frac{\partial^2 l_t}{\partial\beta^*\partial\beta^*}=-\frac{1}{p}\frac{N_1(t)}{(\beta^*)^2}$$
$$\frac{\partial^2 l_t}{\partial\beta^*\partial p}=-\frac{1}{p^2}\frac{N_1(t)}{\beta^*}+\int_0^t\frac{1}{p^2}I_r(s)\big((1-\pi)-\frac{2}{p}\bar{N}_1(s)\big)ds$$
$$\frac{\partial^2 l_t}{\partial\beta^*\partial\pi}=\int_0^t\frac{1}{p}I_r(s)ds$$
$$\frac{\partial^2 l_t}{\partial\pi\partial\pi}=-\sum_{t_i}\frac{1}{p((1-\pi)-\frac{1}{p}N_1(t_{i-}))^2}$$
$$\frac{\partial^2 l_t}{\partial\pi\partial p}=\sum_{t_i}\frac{\pi-1}{(p(\pi-1)-N_1(t_{i-}))^2}-\int_0^t\frac{\beta^*}{p^2}I_r(s)ds$$

$$\begin{aligned}\frac{\partial^2 l_t}{\partial p \partial p} =& -\frac{N_1(t)}{p^2} - \frac{N_1(t)\left(1-\frac{1}{1-p}\right)}{p^2} + \frac{2N_1(t)\left(p+\log(1-p)\right)}{p^3} \\ &+ \frac{1}{p^3}\sum_{t_i}\log\left(\beta^*\frac{1}{p}I_r(t_{i-})\left((1-\pi)-\frac{1}{p}\bar{N}_1(t_{i-})\right)\right) \\ &+ \sum_{t_i}\frac{\frac{I_r(t_{i-})\bar{N}_1(t_{i-})\beta^*}{p^3} - \frac{I_r(t_{i-})\beta^*\left((1-\pi)-\frac{\bar{N}_1(t_{i-})}{p}\right)}{p^2}}{I_r(t_{i-})\beta^*\left((1-\pi)-\frac{\bar{N}_1(t_{i-})}{p}\right)p} \\ &- \frac{1}{p^3}\sum_{t_i}\frac{N_1(t_{i-})\left(3(1-\pi)p-2N_1(t_{i-})\right)}{\left((\pi-1)p+\bar{N}_1(t_{i-})\right)^2} + \frac{2}{p^3}N_1(t) \\ &- \frac{1}{p^4}\int_0^t 2I_r(s)\beta^*\left((1-\pi)p+3\bar{N}_1(s)\right)ds.\end{aligned}$$

The approximate likelihood (and its first and second derivatives with respect to the parameters) becomes involved, especially in terms of $p$ that appears in the denominators of many terms, which makes it no longer possible to use for inference.

## Appendix C Data and Parameters

Table 3 lists the data and parameters used in our analysis, together with information on if we consider them known/unknown or observed/unobserved.

## Appendix D Proof of Unidentifiability

### D.1 Rewriting the ODEs

The key of the proof is to rewrite the set of ODEs of the SIR model with under-reporting in (2) in a more compact form. We also let the initial conditions be as in (3). Recalling that $\beta^* = p\beta_r + (1-p)\beta_u$, and thus $\beta_r I_r(t) + \beta_u I_u(t) = \frac{\beta^*}{p} I_r(t)$, we obtain

$$\begin{aligned}\frac{d}{dt}S(t) =& -\frac{\beta^*}{p}I_r(t)\frac{1}{n}S(t), \\ \frac{d}{dt}I_r(t) =& \beta^* I_r(t)\frac{1}{n}S(t) - \gamma I_r(t),\end{aligned}$$

with initial conditions

$$\begin{aligned}S(0) &= n(1-\pi) - \frac{1}{p}I_r(0), \\ I_r(0) &= ni_0.\end{aligned}$$

**Table 3** Data and parameters used in the analysis.

| Data / Parameter | Description | Notation | Assumed known/unknown |
|---|---|---|---|
| Data | Population size | $n$ | Observed |
| Data | Time step (measured in infectious periods) | $t$ | Observed |
| Data | Initially infectious (reported) | $ni_0$ | Observed |
| Data | Initially infectious (unreported) | $ni_u$ | Unobserved |
| Data | Cumulative reported infections at time $t$ | $N_1(t)$ | Observed |
| Data | Cumulative reported recoveries at time $t$ | $N_2(t)$ | Observed |
| Data | Cumulative unreported infections at time $t$ | $N_3(t)$ | Unobserved |
| Data | Cumulative unreported recoveries at time $t$ | $N_4(t)$ | Unobserved |
| Data | Fraction reported infected during outbreak | $z_r$ | Observed |
| Parameter | Infection rate (reported) | $\beta_r$ | Unknown |
| Parameter | Infection rate (unreported) | $\beta_u$ | Unknown |
| Parameter | Infection rate (overall) | $\beta^* = R_0$ | Unknown |
| Parameter | Reporting fraction | $p$ | Unknown |
| Parameter | Fraction initially immune | $\pi$ | Unknown |
| Parameter | Recovery rate | $\gamma$ | Known |
| Parameter | Exponential growth rate | $\rho$ | Unknown |

Furthermore, we can write the solution for $S(t)$ explicitly as

$$S(t) = S(0) \exp\left\{-\frac{\beta^*}{p}\frac{1}{n}\int_0^t I_r(u)du\right\},$$

which leads to the following ODE for $I_r(t)$

$$\frac{d}{dt}I_r(t) = I_r(t)\left[\beta^*(1-\pi) - \frac{\beta^*}{p}\frac{1}{n}I_r(0)\right]\exp\left\{-\frac{\beta^*}{p}\frac{1}{n}\int_0^t I_r(u)du\right\} - \gamma I_r(t) \tag{12}$$

with initial condition $I_r(0) = ni_0$. Note that the ODE above is enough to define the deterministic epidemic model, since $S(t)$, $I_u(t)$, $R_r(t)$, and $R_u(t)$ can be obtained as

$$S(t) = \left[n(1-\pi) - \frac{1}{p}I_r(0)\right]\exp\left\{-\frac{\beta^*}{p}\frac{1}{n}\int_0^t I_r(u)du\right\} \tag{13}$$

$$I_u(t) = \frac{1-p}{p}I_r(t)$$

$$R_r(t) = p(n - S(t)) - I_r(t)$$

$$R_u(t) = \frac{1-p}{p} R_r(t).$$

### D.2 Two Identical Trajectories of Reported Cases, What can be Said about the Two Sets of Parameters?

Assume that $I_r^1(t)$ and $I_r^2(t)$ solve (12) for two different sets of parameters $(\pi_1, \beta_1^*, p_1)$ and $(\pi_2, \beta_2^*, p_2)$. If the trajectories of reported cases are identical, that is,

$$I_r^1(t) = I_r^2(t), \quad t \geq 0$$

what can be said about two sets of parameters $(\pi_1, \beta_1^*, p_1)$ and $(\pi_2, \beta_2^*, p_2)$?

Let $I_r(t) = I_r^1(t) = I_r^2(t)$. The LHS of (12) is the same for both trajectories, thus the RHS must also be the same (simplifying):

$$\left[\beta_1^*(1-\pi_1) - \frac{\beta_1^*}{p_1}\frac{1}{n}I_r(0)\right]\exp\left\{-\frac{\beta_1^*}{p_1}\frac{1}{n}\int_0^t I_r(u)du\right\} = \left[\beta_2^*(1-\pi_2) - \frac{\beta_2^*}{p_2}\frac{1}{n}I_r(0)\right]\exp\left\{-\frac{\beta_2^*}{p_2}\frac{1}{n}\int_0^t I_r(u)du\right\} \tag{14}$$

Choosing $t = 0$ in (14) above, we obtain

$$\beta_1^*(1-\pi_1) - \frac{\beta_1^*}{p_1}\frac{1}{n}I_r(0) = \beta_2^*(1-\pi_2) - \frac{\beta_2^*}{p_2}\frac{1}{n}I_r(0), \tag{15}$$

which we plug into (14) to obtain

$$\exp\left\{-\frac{\beta_1^*}{p_1}\frac{1}{n}\int_0^t I_r(u)du\right\} = \exp\left\{-\frac{\beta_2^*}{p_2}\frac{1}{n}\int_0^t I_r(u)du\right\}.$$

Thus,

$$\frac{\beta_1^*}{p_1} = \frac{\beta_2^*}{p_2}. \tag{16}$$

Inserting (16) into (15), implies

$$\beta_1^*(1-\pi_1) = \beta_2^*(1-\pi_2). \tag{17}$$

This concludes the proof of the first part of Theorem 1. Furthermore, if one of the parameters is assumed to be known, then the two sets of parameters must be identical because of the derived equations (16) and (17), which proves Corollary 2.

It should also be noted that the identifiable parameter combinations of (16) and (17) possess one degree of freedom. That is, if any single parameter is known a priori, the remaining parameters become structurally identifiable.

In the next section, we prove the second part of Theorem 1.

### D.3 Different Sets of Parameters can lead to Identical Trajectories of Reported Cases

Assume that $I_r^1(t)$ and $I_r^2(t)$ solve (12) for two different sets of parameters $(\pi_1, \beta_1^*, p_1)$ and $(\pi_2, \beta_2^*, p_2)$, with initial condition $I_r^1(0) = I_r^2(0) = ni_0$.

Choose the two sets of parameters $(\pi_1, \beta_1^*, p_1)$ and $(\pi_2, \beta_2^*, p_2)$ so that the conditions (16) and (17) are satisfied. Then, from (12),

$$\begin{aligned} \frac{d}{dt} I_r^1(t) &= I_r^1(t) \left[ \beta_1^*(1-\pi_1) - \frac{\beta_1^*}{p_1} \frac{1}{n} I_r^1(0) \right] \exp\left\{ -\frac{\beta_1^*}{p_1} \frac{1}{n} \int_0^t I_r^1(u) du \right\} - \gamma I_r^1(t) \\ &= I_r^1(t) \left[ \beta_2^*(1-\pi_2) - \frac{\beta_2^*}{p_2} \frac{1}{n} I_r^1(0) \right] \exp\left\{ -\frac{\beta_2^*}{p_2} \frac{1}{n} \int_0^t I_r^1(u) du \right\} - \gamma I_r^1(t), \end{aligned}$$

that is, the ODEs for $I_r^1(t)$ and $I_r^2(t)$ are identical. Thus, we have shown that two different sets of parameters satisfying conditions (16) and (17) lead to two identical trajectories of the reported cases, $I_r^1(t) = I_r^2(t),\ t \geq 0$.

Note that the other trajectories $S^1(t), I_u^1(t), R_r^1(t), R_u^1(t)$ and $S^2(t), I_u^2(t), R_r^2(t), R_u^2(t)$ which can be obtained through (13), are instead *not* identical.

This concludes the proof of the second, and final, part of Theorem 1.

**Acknowledgements** We thank Gianpaolo Scalia Tomba for helpful discussions and verification of the simulation code.

**Author Contributions** **Fanny Bergström:** Conceptualization, Methodology, Software, Formal analysis, Data Curation, Writing - Original Draft, Writing - Review & Editing, Visualization. **Martina Favero:** Conceptualization, Methodology, Validation, Formal analysis, Writing - Review & Editing. **Tom Britton:** Conceptualization, Methodology, Writing - Review & Editing, Supervision.

**Funding** Open access funding provided by Stockholm University. M. F. acknowledges the Knut and Alice Wallenberg Foundation (Program for Mathematics, grant 2020.072) for financial support. T.B. acknowledges financial support from the Swedish Research Council (grant 2020-0474).

**Data Availability** The R and Julia code and data for the simulations and results are publicly available on GitHub: https://github.com/fannybergstrom/identifiability.

## Declarations

**Conflicts of interest** The authors declare that they have no known competing financial interests or personal relationships that could have appeared to influence the work reported in this paper.